# Simple exercises that significantly increase cerebral blood flow and cerebral oxygenation.

**Alexander Gersten[1,*], Jacqueline Perle[1], Amir Raz[2] and Robert Fried[1]**

[1]Department of Psychology, Hunter College of the City University of New York, NY, USA

[2]Department of Psychiatry, Columbia University College of Physicians & Surgeons and New York State Psychiatric Institute, NY, USA

Abstract

We tested the hypothesis that simple exercises may significantly increase cerebral blood flow (CBF) and/or cerebral oxygenation. Eighteen subjects ranging in age from nineteen to thirty nine participated in a four-stage study during which measurements of end tidal $CO_2$ (EtCO2 - by capnometer) and local brain oxygenation (by near-infrared spectroscopy (NIRS) sensor) were taken. The four stages were 1) baseline, 2) breathing exercises, 3) solving an arithmetic problem, and 4) biofeedback. During the breathing exercises there was a significant increase in $EtCO_2$ indicating a significant increase in global CBF. The increase in global CBF was estimated on the basis of a theoretical model. During the arithmetic and biofeedback tasks there was a significant increase in the local (Fp1) oxygenation, but it varied between the different participants. The results may lead to new clinical applications of CBF and brain oxygenation monitoring and behavioral control. We foresee future more detailed investigations in the control of $CO_2$ in brain circulation in specific regions of the brain involved in cognition and memory.

[*]On sabbatical leave from the Department of Physics, Ben-Gurion University of the Negev, Beer-Sheva, Israel.

# Introduction

Can simple exercises be devised to increase cerebral blood flow (CBF) and/or cerebral oxygenation? We investigated exactly that question by using three different techniques, namely: a simple breathing procedure, solving an arithmetic problem and biofeedback.

Elsewhere (Gersten at al., 2007) we have analyzed the influence of arterial partial pressure of $CO_2$ ($PaCO_2$) on CBF and found that it may dramatically change the CBF. The changes involve the blood flow of the whole brain. It is a global effect. These results were used in another investigation (Gersten et al., 2007b) in which yoga practitioners were increasing their $PaCO_2$ through periodic yoga (pranayama) breathing techniques.

We will demonstrate that significant increase of $PaCO_2$ (and of total CBF) can be achieved with untrained people using very simple breathing procedures. The reason for that is the dependence of $PaCO_2$ on ventilation (West, 1992)

$$PaCO_2 = K\frac{\dot{V}_{CO_2}}{\dot{V}_A}, \ \ K = 863\,mm\,Hg \tag{1}$$

where $\dot{V}_{CO_2}$ is the $CO_2$ production (dependent on the metabolism) and $\dot{V}_A$ is the alveolar ventilation. This means that the $PaCO_2$ is inversely proportional to ventilation. Therefore it is possible to control the $PaCO_2$ by either breathing slowly (and not increasing the tidal volume substantially) or by holding the breath. Untrained people increase their tidal volume while breathing more slowly, but the overall effect is usually a slight increase in $PaCO_2$. People trained in breathing exercises may increase their $PaCO_2$ considerably by learning to control their tidal volume. For very small ventilation a correction is needed to the $PaCO_2$ formula (Riggs, 1970)

$$PaCO_2 = K \frac{\dot{V}_{CO_2}}{\dot{V}_A - \dot{V}_D}, \quad \dot{V}_D = 2.07 \ l/\min \quad , \tag{2}$$

where $\dot{V}_D$ is the contribution of the dead space.

It is well known that concentrating on a mental problem changes brain's oxygenation locally (Chance et al, 1993). This finding was also applied further to solving a simple arithmetic problem on local brain oxygenation near the Fp1 area.

Hershel Toomim (Toomim et al., 2004) developed the device called hemoencephalograph (HEG), whose readings is related to regional cerebral oxygenation (Gersten et. al, 2007c). The device has many advantageous features which allowed us to use it in our experiment. Toomim has observed that he can influence the results by looking at the HEG display, which is essentially a biofeedback technique used by us as well. As a result of Toomim findings many biofeedback experiments were conducted with the HEG, confirming the effect. The readings of the HEG are very sensitive to changes in the range of normal oxygenation of the brain. This is not the case with INVOS brain oximeters used in operation rooms whose main aim is to detect abnormally low oxygenation states. For that reason we preferred to use the HEG to detect biofeedback effects even though it is much simpler and less sophisticated compared with INVOS cerebral oximeters (Gersten et al, 2007c).

The readings of the HEG are normalized to 100 (SD=20), the average on 154 adult attendants at professional meetings (Toomim et al, 2004). We have compared (Gersten et al, 2007c) the readings of HEG with the regional saturation of oxygen (rSO$_2$) readings of the INVOS cerebral oximeter of Somanetics. This allowed us to make estimates of the ratios of rSO$_2$ using the HEG. We found

$$x_1/x_2 = \log(y_1/32.08)/\log(y_2/32.08), \quad x \equiv rSO_2, \quad y \equiv HEG \text{ readings.} \qquad (3)$$

Measurements were taken using HEG and a capnometer (a device measuring end tidal $CO_2$) simultaneously. End tidal $CO_2$ is closely related to $PaCO_2$. Eighteen subjects participated in the experiment in which HEG and $CO_2$ data were recorded for 5 intervals of baseline, simple breathing exercises, simple arithmetic tasks and biofeedback. The results show that almost all participants could increase their brain oxygenation or CBF, but in each case it was strongly dependent on one of the three methods used. We can conclude that it is possible to substantially increase local oxygenation or global CBF using one of the three methods described above, but the preferred method is highly individual.

The protocol of this research was approved by the IRB of Hunter College of the City University of New York.

## METHODS AND MATERIALS

It is well known that breathing patterns affect the $CO_2$ levels in the arteries (Fried and Grimaldi, 1993), which in turn can affect cerebral (brain's) blood circulation and oxygenation. Mental work and biofeedback may affect both local as well as global oxygen levels in the brain.

The influence of breathing exercises, problem solving and biofeedback on brain oxygen and CO2 arterial levels were considered in an experiment outlined below. The experiment dealt with 3 topics

1. The physiological effects of mild breathing exercises on increasing $CO_2$ and oxygen levels in the brain.

2. The physiological effects of problem solving (a particular case of mental performance) on the $CO_2$ and oxygen levels in the brain.

3. The physiological effects of biofeedback on the $CO_2$ and oxygen levels in the brain.

An experiment dealing with the second and third topic gives qualitative information about how much the oxygen levels will rise during problem solving and biofeedback, while an experiment dealing with the first topic will give the same information, but this time, from breathing exercises.

It is important to note that in the first topic global CBF is concerned, while in the second and third topic the local brain oxygenation at the Fp1 area.

The participants were connected to the two devices needed for the experiment: the capnometer that measured end tidal $CO_2$ and the Cerebral Oximeter. The connection used was made via a sensor placed on the forehead at Fp1.

The $CO_2$ levels were estimated using a capnometer produced by Better Physiology LTD, which measures end tidal $CO_2$ ($EtCO_2$) of the exhaled air ($EtCO_2$ is highly correlated with the $PaCO_2$ levels of the arteries). All participants received their own new nasal insert which were sterilized before each use and connected to the capnometer. The data were detected via USB output cable connected to the computer and stored for subsequent review.

The oxygen levels were estimated using two devices: the HEG which was calibrated to the INVOS Cerebral Oximeter produced by Somanetics Corp. (see www.somanetics.com) and based on advanced near infrared spectroscopy (NIRS) technology. A sensor was attached to the forehead measuring the oxygenation in a depth of about one inch inside the brain. The devices that were used were non-invasive and FDA approved, fully automated and did not require special precautions. The data were stored on a computer.

The data of the capnometer and oximeters were combined together and analyzed using Matlab subprograms.

The participants were asked to do paced breathing exercises as instructed by the experimenters.

Before the 3 experiments baseline data were taken for 5 minutes using the INVOS oximeter, and another 5 minutes using the HEG and capnometer. The participants were prevented from seeing the screens of the devices in order to avoid biofeedback.

In the first experiment (lasting 5 minutes) participants were asked to walk slowly, breathe in for 3 steps, hold their breath during the next 3 steps, exhale during the next 3 steps, and hold their breath for the next 3 steps after exhaling. We made sure that the participants understood these instructions. The participants also did not see the screens of the devices in order to avoid biofeedback.

In the second experiment (lasting 5 minutes) participants were given an arithmetical problem to solve while being attached to the HEG and capnometer. The theoretical basis for this experiment is that more oxygen is needed while solving problems. A simple arithmetical problem of subtracting the number 7 continuously, starting from 1200 (1193,1186,...) was used. The participants were again prevented to see the screens of the devices in order to avoid biofeedback.

In the third experiment (lasting 5 minutes) participants were asked to look at the HEG display trying to raise the curve by mental feedback. This time they were allowed to look at the display.

## Participants

The participants were 18 participants from the introductory course to psychology (PSY 100) in Hunter College of the City University of New York.

 All participants had to sign an informed consent.

At least two experimenters were present during each experiment.

The confidentiality of the participants was protected.

## Results

To better illustrate the results a few examples of HEG  and $CO_2$  data will be included.

In Fig. 1, subplot $HC3_1$, the baseline data of participant No. 3 are displayed. The HEG baseline was not constant during the 5 minutes of data taking.

In the subplot $HC3_2$ the result of the breathing exercise are displayed. Even though this was a first trial, the $CO_2$ pattern seems to be quite periodic. The $CO_2$ pattern has a periodicity of about 15 secs per period, while normal breathing has a periodicity of about 4 secs per period. The prolongation of the respiratory period should lead to an accumulation of arterial $CO_2$ and an increase of global CBF, provided there is no greater increase in the tidal volume. In this case there was only small increase in arterial $CO_2$ but a significant increase in oxygenation (HEG). The increase of arterial $CO_2$ depends on the control over the tidal volume. Individuals trained in this breathing exercise, can easily increase their arterial $CO_2$ by about 20-30%.

Solving the arithmetic problem (see Fig. 1, subplot $HC3_3$) led to an increase of the HEG readings (oxygenation), but we notice that the respiration was speeded up probably due to increased tension. After all, to subtract 7 and calculate and evaluate the result in the mind is not a very pleasant enterprise.

Subplot $HC3_4$ of Fig. 1 is very interesting. In this case the subject was looking at the display of the HEG line trying mentally to raise it up. The performance (without previous training) is very impressive. Starting from baseline values the HEG readings were climbing up for about 2.5 minutes to values about 30% higher (a local increase of oxygenation at the Fp1 area). This case teaches us that in the evaluation of the results we must not only consider average values but also the maximal values which are an indicative of the possible potential. The respiration pattern indicates a slow but a very deep breathing (hyperventilation). The arterial $CO_2$ decreased for more than 20% which should lead to a significant decrease of global CBF. The biofeedback was very successful in spite of the hyperventilation.

In Table 1 the HEG mean values of the four cases (baseline, breathing exercise, arithmetic problem and biofeedback) are given for all 18 participants. While analyzing the data we must take into account that the participants were performing their tasks for the first time. Most of the tasks were performed relatively well. As shown in table 1 most participants were able to increase their mean HEG readings in at least one task. When averaging all participant's data there were no significant changes were found, indicating that no one method was preferable.

In this respect table 2 is more informative, it gives the ratio of maxima to the mean of the baseline. The maxima indicate the potential of the exercises. These maximal values should be easily reached with practice. There was a maximal increase of 30% during breathing exercises, 32% during solving the arithmetic problem and 28% during the biofeedback.

Table 2a displays the same results as Table 2 but with $rSO_2$ ratios determined according to Eq. (3). The results are quite similar, indicating that the HEG ratios are quite reliable in estimating the $rSO_2$ changes.

Of the 18 participants 14 were able to increase the HEG readings by at least 10% during one of the exercises (5 during the breathing exercise, 9 while solving the arithmetic problem, 8 during the biofeedback).

Of the 18 participants 7 were able to increase the HEG readings by at least 18% during one of the exercises (3 during the breathing exercise, 6 while solving the arithmetic problem, 5 during the biofeedback).

The breathing exercise was the most difficult for the participants. It took them an average of 6 minutes to fully understand the instructions. The exercise required some discipline and experience. Most of the participants performed it relatively well.

Fig. 2 shows that participant No. 9 has performed the breathing exercises relatively well (subplot $HC9_2$). His $CO_2$ pattern was periodic and amplitude stable. The pattern was not completely smooth. This is understandable, since it was his first attempt to perform the exercise. In the same subplot, the corresponding HEG curve is very interesting. The HEG waveform has the same period as the $CO_2$ pattern. In Fig. 3 the spectral analyses of the subplots of Fig. 2 is presented. Subplots $HC9_5$, $HC9_6$, $HC9_7$, $HC9_8$, correspond to subplots $HC9_1$, $HC9_2$, $HC9_3$, $HC9_4$, respectively. Only for the breathing exercise (subplot $HC9_6$) there was a correlation between the $CO_2$ pattern and the HEG waveform. The maxima of the corresponding spectral powers exactly coincided at 0.053 Hz. There was no correlation between the HEG and $CO_2$ patterns in the baseline (subplot $HC9_5$), while solving the arithmetic problem (subplot $HC9_7$) and during the biofeedback (subplot $HC9_8$).

In Fig. 4 participant No. 2 had difficulty performing the breathing exercise (the $CO_2$ pattern in subplot $HC2_2$). Although the 3 step pattern was kept correctly, the participant was inhaling during some of the breath holding periods. When done correctly the 3 step breathing cycle should last for about 15 seconds. As the participant was breathing in between, the average breath length was only 7.2 seconds (see Table 3). Interestingly the HEG waveform of subplot $HC2_2$ has a period of about 15 seconds irrespective of the breathing in between the 3 step pattern.

The performance of the breathing exercise for all participants is summarized in Table 3 and Table 4. In table 3 the average breath length is given. When well performed, it should be longer than 10 seconds. Twelve of 18 participants have performed the breathing exercise well.

Interestingly the HEG waveform has the same periodicity as the $CO_2$ pattern. This can be seen in table 4, where the maxima of the power spectra of the $CO_2$ pattern and the HEG waveform are displayed. Here in 15 out of 18 cases there is a coincidence of the maxima position. This coincidence is well presented in Fig. 5, where the correlation between the power spectra of the $EtCO_2$ periodic pattern and the corresponding HEG periodic pattern is depicted by their multiplication. The power spectra are normalized to unity. Maximal correlation is achieved when the multiplication is equal to 1.

Additional examples of the correlation between the $CO_2$ pattern and the HEG waveform are given in Figs. 6-11.

In Table 5 the average values of the end tidal $CO_2$ are given for the four cases: baseline, breathing exercise, while solving the arithmetic problem and for the biofeedback. The last column lists the percent increase in the breathing exercises compared to baseline. We note that the simple breathing exercises led to a substantial increase of $PaCO_2$ and indirectly to a significant increase of the global CBF. Estimates of CBF can be found in Table 6. They are based on Eq. 5.3 and Table 1 of the chapter "Peculiarities of cerebral blood flow, the role of carbon dioxide" (Gersten et al, 2007a). Comparing the data of the breathing exercises in Table 1 with Table 5, in spite of the large increases of end tidal $CO_2$ and expected large increases of global CBF, we did not find a corresponding similar increase with the readings of the HEG at Fp1.

## Conclusions

Our method to obtain the above results is through the use of human subjects. This is a new avenue in approaching the study of CBF, brain oxygenation, improving the cognitive function and especially in view of the growing elderly population.

Our three methods are simple, can be used on the general population, are non-invasive, without the use pharmaceuticals and have no side effects. They differ from each other in that the breathing affects mostly the global blood flow, arithmetic problem solving and biofeedback affects the regional blood flow (the Fp1 region).

Both our theoretical and experimental work differs from other studies due the specific instrumentation and our experimental procedure. Most of the results came close to our expectation.

We concluded that breathing can be used effectively to control CBF by the ventilatory control of end tidal $CO_2$. This research may have implications for complementary diagnosis and treatment of conditions involving regional cerebral metabolism such as cerebral vascular ischemia, seizures disorders, stroke, Alzheimer's disease, and more. Following that thought could lead us to improved cognitive function through a higher supply of oxygen to specific regions of the brain.

We foresee future more detailed investigations to be made in the area of the effect of $CO_2$ on specific regions of the brain. This would be of great interest because a higher $CO_2$ supply results in a higher blood flow and thus to more oxygen and better overall brain function, specifically cognitive function.

**Table 1.**  Mean values of HEG readings (with standard deviations in parenthesis) for the four cases: baseline, breathing exercise, arithmetic problem and biofeedback.

| N | Baseline | Breath.Ex. | Arith.Prob. | Biofeedback |
|---|---|---|---|---|
| 1 | 69.3(2.1) | 75.5(6.5) | 83.0(4.2) | 77.8(2.2) |
| 2 | 100.1(0.8) | 100.3(7.9) | 97.2(2.0) | 91.5(1.8) |
| 3 | 108.7(2.4) | 116.9(1.8) | 113.3(1.1) | 128.7(9.4) |
| 4 | 89.7(1.6) | 78.7(1.3) | 94.3(1.5) | 95.4(3.7) |
| 5 | 158.9(2.9) | 165.0(2.2) | 178.5(4.7) | 155.9(2.3) |
| 6 | 70.7(3.3) | 65.6(2.1) | 62.5(1.4) | 62.5(1.4) |
| 7 | 96.3(1.7) | 94.6(2.0) | 113.9(3.7) | 110.6(5.9) |
| 8 | 119.2(1.6) | 114.7(0.9) | 121.1(1.3) | 118.8(1.1) |
| 9 | 99.3(4.4) | 88.3(3.5) | 103.9(3.2) | 98.2 (10) |
| 10 | 92.9(2.6) | 93.7(6.3) | 93.9(0.8) | 89.8(2.4) |
| 11 | 94.1(1.1) | 96.4(2.2) | 97.0(2.5) | 96.2(2.2) |
| 12 | 120.1(5.2) | 108.2(8.5) | 138.7(9.5) | 131.6(6.5) |
| 13 | 158.4(5.2) | 133.9(9.8) | 177.4(13.5) | 150.6(5.4) |
| 14 | 110.3(1.6) | 99.1(1.8) | 103.4(1.5) | 100.5(1.2) |
| 15 | 104.0(2.5) | 93.8(7.5) | 121.6(4.5) | 121.6(4.5) |
| 16 | 90.9(1.6) | 87.8(3.2) | 101.3(1.8) | 88.5(1.7) |
| 17 | 63.0(1.8) | 60.3(1.7) | 62.4(1.3) | 60.0(1.5) |
| 18 | 97.6(3.9) | 95.6(2.5) | 113.0(3.9) | 106.4(2.1) |
| mean(SD) | 102.4(25.7) | 98.3(24.4) | 109.8(31.3) | 104.7(26.5) |

**Table 2.** Ratios of the maxima of the HEG to the mean of the baseline

| N | Baseline | Breath.Ex. | Arith.Prob. | Biofeedback |
|---|---|---|---|---|
| 1 | 1.09 | 1.30 | 1.32 | 1.20 |
| 2 | 1.02 | 1.18 | 1.05 | 0.96 |
| 3 | 1.06 | 1.11 | 1.08 | 1.28 |
| 4 | 1.09 | 0.91 | 1.09 | 1.14 |
| 5 | 1.03 | 1.07 | 1.17 | 1.02 |
| 6 | 1.08 | 0.98 | 0.94 | 0.94 |
| 7 | 1.04 | 1.03 | 1.25 | 1.23 |
| 8 | 1.03 | 0.98 | 1.04 | 1.03 |
| 9 | 1.09 | 0.99 | 1.11 | 1.14 |
| 10 | 1.07 | 1.19 | 1.03 | 1.04 |
| 11 | 1.03 | 1.10 | 1.08 | 1.07 |
| 12 | 1.09 | 1.07 | 1.28 | 1.18 |
| 13 | 1.07 | 0.99 | 1.26 | 1.02 |
| 14 | 1.04 | 0.94 | 0.98 | 0.94 |
| 15 | 1.07 | 1.04 | 1.26 | 1.26 |
| 16 | 1.03 | 1.05 | 1.16 | 1.00 |
| 17 | 1.07 | 1.03 | 1.05 | 1.02 |
| 18 | 1.05 | 1.04 | 1.20 | 1.14 |
| mean(SD) | 1.06(.02) | 1.06(.10) | 1.13(.11) | 1.09(.11) |

**Table 2a.** Ratios of the maxima of $rSO_2$ to the mean of the baseline as determined from HEG readings via Eq. (3).

| N | Baseline | Breath.Ex. | Arith.Prob. | Biofeedback |
|---|----------|------------|-------------|-------------|
| 1 | 1.11 | 1.34 | 1.36 | 1.24 |
| 2 | 1.01 | 1.15 | 1.04 | 0.97 |
| 3 | 1.05 | 1.09 | 1.07 | 1.20 |
| 4 | 1.08 | 0.91 | 1.08 | 1.13 |
| 5 | 1.02 | 1.04 | 1.10 | 1.01 |
| 6 | 1.09 | 0.98 | 0.92 | 0.92 |
| 7 | 1.04 | 1.03 | 1.20 | 1.19 |
| 8 | 1.02 | 0.98 | 1.03 | 1.02 |
| 9 | 1.08 | 0.99 | 1.10 | 1.12 |
| 10 | 1.06 | 1.17 | 1.03 | 1.03 |
| 11 | 1.03 | 1.09 | 1.07 | 1.06 |
| 12 | 1.06 | 1.05 | 1.19 | 1.12 |
| 13 | 1.04 | 0.99 | 1.15 | 1.01 |
| 14 | 1.03 | 0.95 | 0.99 | 0.95 |
| 15 | 1.06 | 1.04 | 1.19 | 1.19 |
| 16 | 1.03 | 1.05 | 1.14 | 1.00 |
| 17 | 1.11 | 1.04 | 1.08 | 1.03 |
| 18 | 1.04 | 1.04 | 1.16 | 1.12 |

**Table 3.** The average breath length in seconds

| N | Baseline | Breath.Ex. | Arith.Prob. | Biofeedback |
|---|---|---|---|---|
| 1 | 3.1 | 11.2 | 2.9 | 3.5 |
| 2 | 3.1 | 7.2 | 3.7 | 7.8 |
| 3 | 5.5 | 10.8 | 3.2 | 8.8 |
| 4 | 4.6 | 15.0 | 3.3 | 4.6 |
| 5 | 3.1 | 6.0 | 3.1 | 3.1 |
| 6 | 7.7 | 14.8 | 7.7 | 8.4 |
| 7 | 5.1 | 11.7 | 5.1 | 5.1 |
| 8 | 5.9 | 15.2 | 5.0 | 9.6 |
| 9 | 6.3 | 16.1 | 5.3 | 7.6 |
| 10 | 2.9 | 5.6 | 3.0 | 4.1 |
| 11 | 3.0 | 7.8 | 3.0 | 3.5 |
| 12 | 2.7 | 8.1 | 2.5 | 9.3 |
| 13 | 3.6 | 15.3 | 3.1 | 4.7 |
| 14 | 2.9 | 4.7 | 3.0 | 3.1 |
| 15 | 4.7 | 10.3 | 4.8 | 5.1 |
| 16 | 4.2 | 11.3 | 3.7 | 7.0 |
| 17 | 4.5 | 11.6 | 3.1 | 5.0 |
| 18 | 4.2 | 15.8 | 4.0 | 11.0 |
| mean(SD) | 4.3(1.4) | 11.0(3.8) | 3.9(1.3) | 6.2(2.5) |

**Table 4**. The spectral analysis (breathing exercise) positions of the maxima in Hz. Last column is the Pearson correlation coefficient r.

| N | EtCO$_2$ | HEG | r |
|---|---|---|---|
| 1 | 0.0579 | 0.0580 | 0.6881 |
| 2 | 0.0714 | 0.0709 | 0.8934 |
| 3 | 0.0677 | 0.0670 | 0.8127 |
| 4 | 0.0621 | 0.0624 | 0.9737 |
| 5 | 0.1194 | 0.0818 | -0.2721 |
| 6 | 0.0667 | 0.0642 | 0.5778 |
| 7 | 0.0681 | 0.0731 | 0.7778 |
| 8 | 0.0531 | 0.0554 | 0.8076 |
| 9 | 0.0527 | 0.0529 | 0.9398 |
| 10 | 0.0686 | 0.0651 | 0.3986 |
| 11 | 0.1118 | 0.0556 | -0.0638 |
| 12 | 0.0512 | 0.0513 | 0.9422 |
| 13 | 0.0656 | 0.0658 | 0.9521 |
| 14 | 0.1217 | 0.0819 | 0.2478 |
| 15 | 0.0643 | 0.0642 | 0.9095 |
| 16 | 0.0816 | 0.0830 | 0.8192 |
| 17 | 0.0594 | 0.0597 | 0.9717 |
| 18 | 0.0531 | 0.0570 | 0.4862 |

**Table 5.** Mean values of EtCO$_2$. Last column is the increase in % due to breathing exercises compared to baseline.

| N | Baseline | Breath.Ex. | Arith.Prob. | Biofeedback | % increase |
|---|----------|------------|-------------|-------------|------------|
| 1 | 37.5 | 46.9 | 39.4 | 34.5 | 25.2 |
| 2 | 36.6 | 44.9 | 39.2 | 34.4 | 22.8 |
| 3 | 40.1 | 40.7 | 42.1 | 34.0 | 1.5 |
| 4 | 35.4 | 38.4 | 38.0 | 32.0 | 8.3 |
| 5 | 37.5 | 36.7 | 38.5 | 39.1 | -2.0 |
| 6 | 36.9 | 39.6 | 35.6 | 36.3 | 7.4 |
| 7 | 31.4 | 39.8 | 38.3 | 38.3 | 27.0 |
| 8 | 37.5 | 46.2 | 40.2 | 38.5 | 23.2 |
| 9 | 37.5 | 41.9 | 35.5 | 34.6 | 11.7 |
| 10 | 33.3 | 38.8 | 36.4 | 36.4 | 16.5 |
| 11 | 37.1 | 39.6 | 37.5 | 38.6 | 6.5 |
| 12 | 40.0 | 43.7 | 40.3 | 40.6 | 9.1 |
| 13 | 38.2 | 45.9 | 39.3 | 38.7 | 20.1 |
| 14 | 33.3 | 34.1 | 35.9 | 36.3 | 2.4 |
| 15 | 38.2 | 45.7 | 42.3 | 40.6 | 19.8 |
| 16 | 28.6 | 33.9 | 28.9 | 28.8 | 18.4 |
| 17 | 40.3 | 43.3 | 41.2 | 39.0 | 7.6 |
| 18 | 35.4 | 42.6 | 40.6 | 37.9 | 20.3 |
| mean(SD) | 36.4(3.1) | 41.3(4.0) | 38.3(3.2) | 36.6(3.1) | 13.7(8.8) |

**Table 5a.** Ratios of the maxima of EtCO$_2$ to the mean of the baseline

| N | Baseline | Breath.Ex. | Arith.Prob. | Biofeedback |
|---|----------|------------|-------------|-------------|
| 1 | 1.05 | 1.39 | 1.09 | 1.20 |
| 2 | 1.03 | 1.45 | 1.13 | 1.10 |
| 3 | 1.06 | 1.11 | 1.11 | 1.07 |
| 4 | 1.05 | 1.14 | 1.11 | 1.01 |
| 5 | 1.05 | 1.08 | 1.12 | 1.13 |
| 6 | 1.06 | 1.22 | 0.96 | 1.09 |
| 7 | 1.09 | 1.43 | 1.44 | 1.44 |
| 8 | 1.15 | 1.35 | 1.15 | 1.11 |
| 9 | 1.09 | 1.21 | 1.07 | 1.14 |
| 10 | 1.08 | 1.38 | 1.16 | 1.25 |
| 11 | 1.09 | 1.19 | 1.07 | 1.12 |
| 12 | 1.04 | 1.35 | 1.12 | 1.14 |
| 13 | 1.08 | 1.29 | 1.12 | 1.11 |
| 14 | 1.05 | 1.13 | 1.12 | 1.18 |
| 15 | 1.12 | 1.42 | 1.21 | 1.22 |
| 16 | 1.05 | 1.31 | 1.14 | 1.06 |
| 17 | 1.08 | 1.23 | 1.08 | 1.08 |
| 18 | 1.07 | 1.32 | 1.26 | 1.25 |

**Table 6.** Expected global CBF in ml/100g/min. Last column is the expected increase in % compared to baseline due to breathing exercises.

| N | Baseline | Breath.Ex. | % increase |
|---|----------|------------|------------|
| 1 | 44.9±1.8 | 67.0±3.5 | 49.2±9.9 |
| 2 | 43.7±1.8 | 60.3±2.9 | 38.1±8.6 |
| 3 | 49.1±2.0 | 50.2±2.1 | 2.2±6.0 |
| 4 | 42.3±1.7 | 46.2±1.9 | 9.4±6.3 |
| 5 | 44.9±1.8 | 43.9±1.8 | -2.3±5.6 |
| 6 | 44.1±1.8 | 48.2±2.0 | 9.5±6.3 |
| 7 | 38.3±1.6 | 48.6±2.0 | 27.2±7.6 |
| 8 | 45.0±1.8 | 64.6±3.3 | 43.6±9.3 |
| 9 | 44.9±1.8 | 52.7±2.3 | 17.3±6.9 |
| 10 | 40.0±1.7 | 46.9±1.9 | 17.3±6.9 |
| 11 | 44.4±1.8 | 48.1±2.0 | 8.3±6.3 |
| 12 | 49.0±2.0 | 56.9±2.6 | 16.2±7.1 |
| 13 | 46.0±1.9 | 63.4±3.2 | 38.0±8.9 |
| 14 | 39.9±1.7 | 40.7±1.7 | 2.0±6.0 |
| 15 | 45.9±1.9 | 62.9±3.1 | 37.0±8.8 |
| 16 | 36.2±1.6 | 40.6±1.7 | 12.2±6.9 |
| 17 | 49.4±2.0 | 56.1±2.5 | 13.4±6.9 |
| 18 | 42.2±1.7 | 54.2±2.4 | 28.4±7.7 |

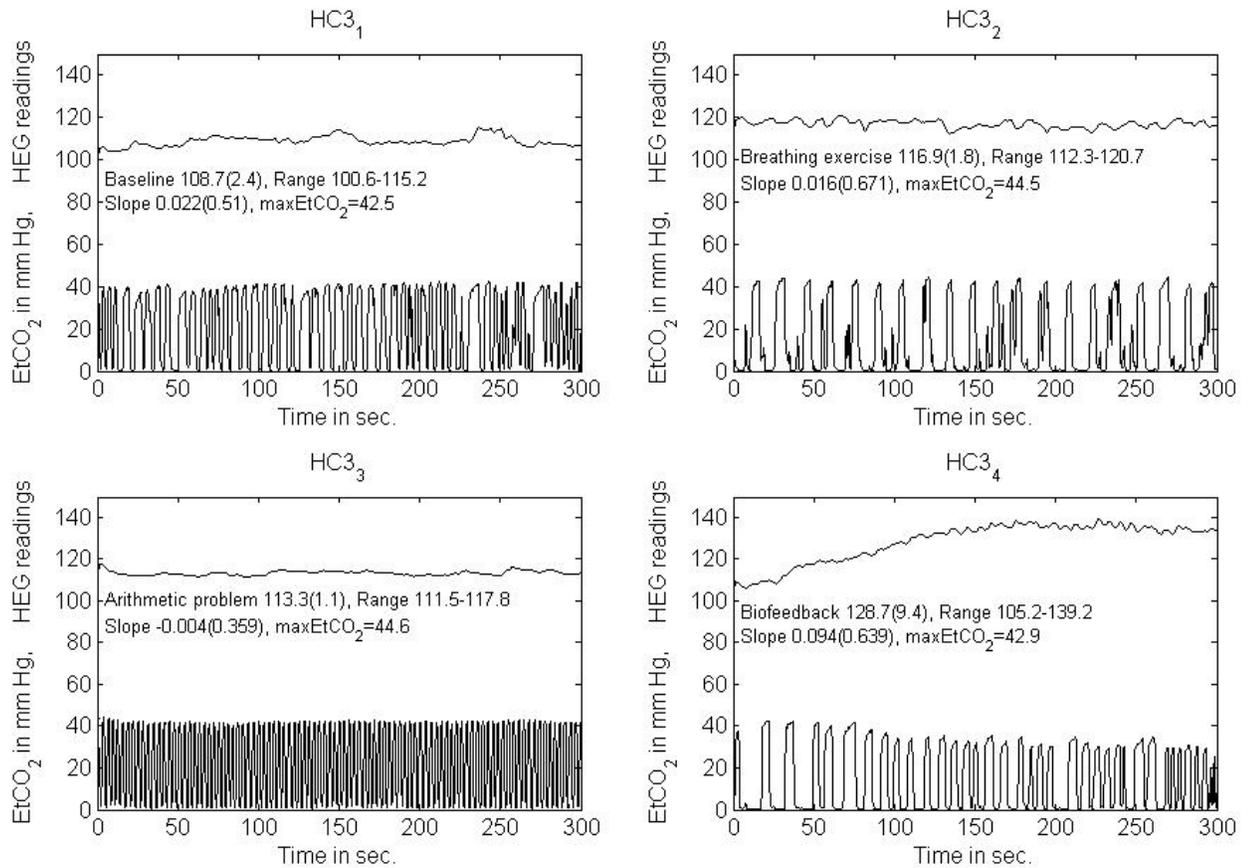

**Fig.1,** The HEG readings and the capnometer displays (EtCO$_2$) are shown for the 4 cases: baseline (subplot HC3$_1$), breathing exercise (subplot HC3$_2$), arithmetic problem (subplot HC3$_3$) and biofeedback (subplot HC3$_4$). The upper curves are the HEG readings, the lower curves the CO$_2$ values (at exhale). The numbers with numbers in parenthesis are the mean values and standard deviations.

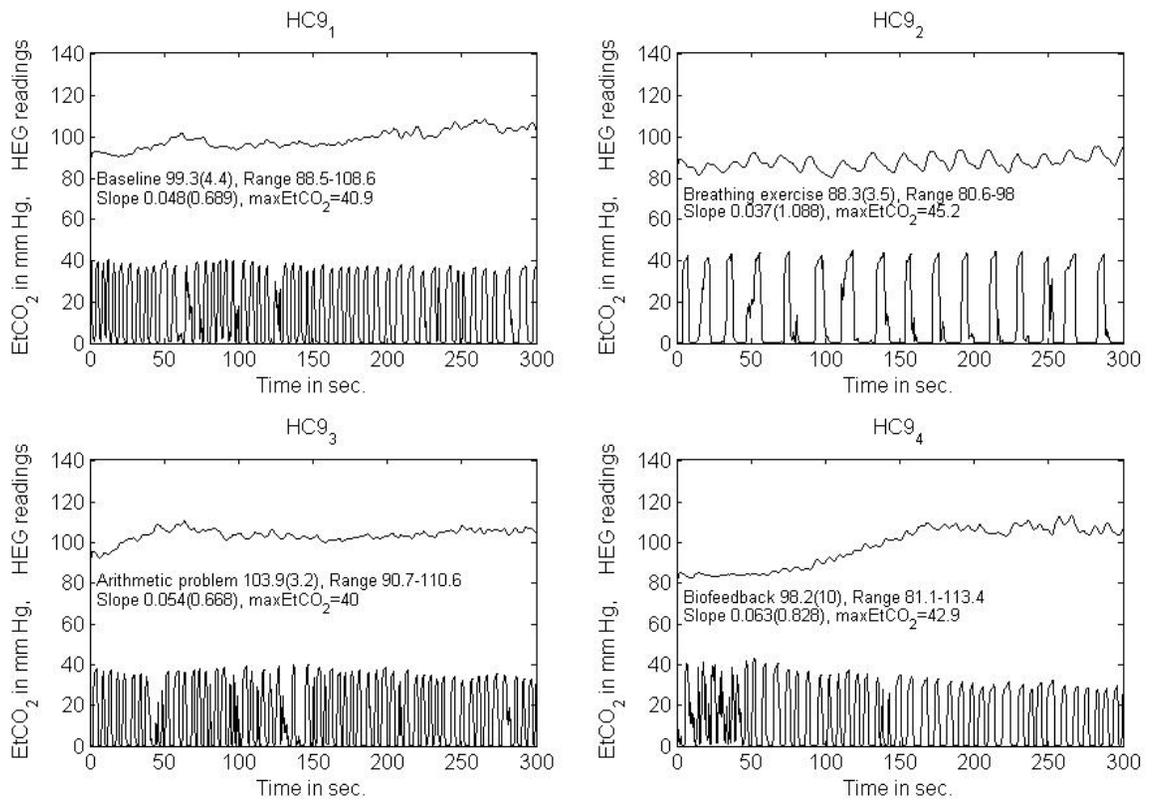

**Fig.2,** The HEG readings and the capnometer displays (EtCO$_2$) are shown for the 4 cases: baseline (subplot HC9$_1$), breathing exercise (subplot HC9$_2$), arithmetic problem (subplot HC9$_3$) and biofeedback (subplot HC9$_4$). The upper curves are the HEG readings, the lower curves the CO$_2$ values (at exhale).

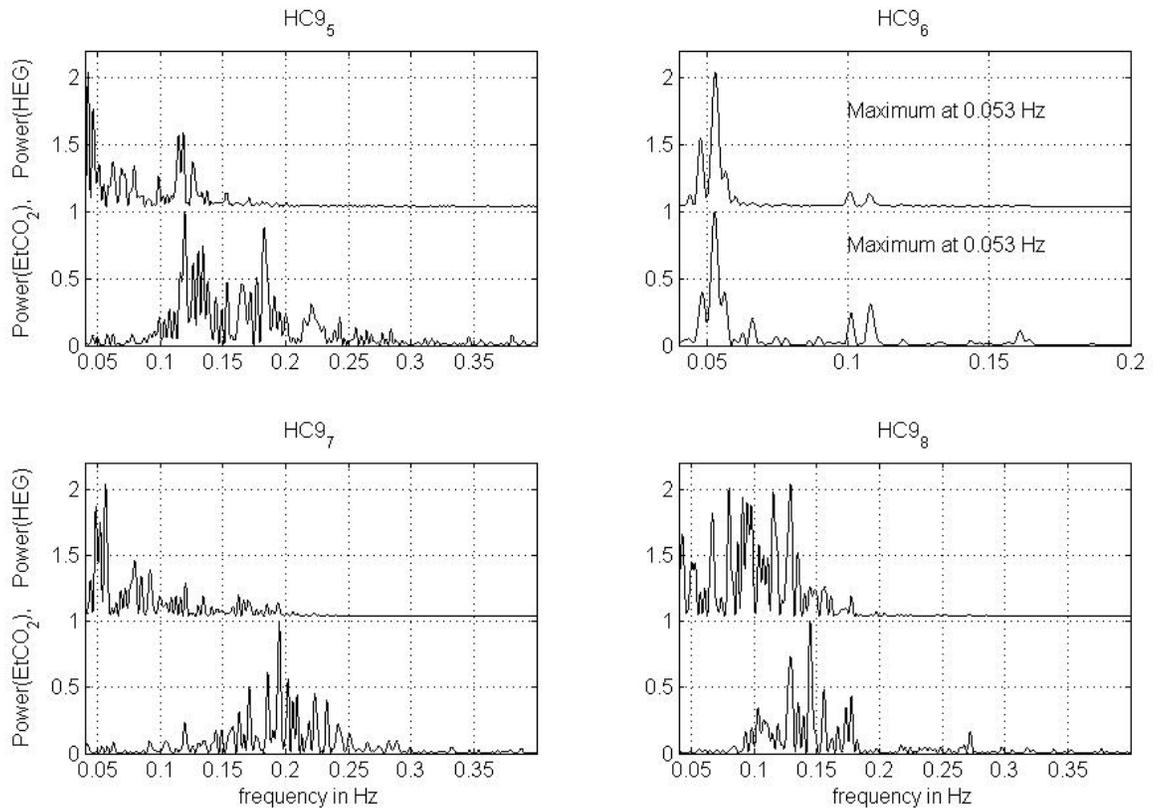

**Fig.3,** Spectral analyses of the HEG readings and the capnometer displays (EtCO$_2$) are shown for the 4 cases: baseline (subplot HC9$_5$), breathing exercise (subplot HC9$_6$), arithmetic problem (subplot HC9$_7$) and biofeedback (subplot HC9$_8$). The upper curves are for the HEG readings, the lower curves for the CO$_2$ values (at exhale).

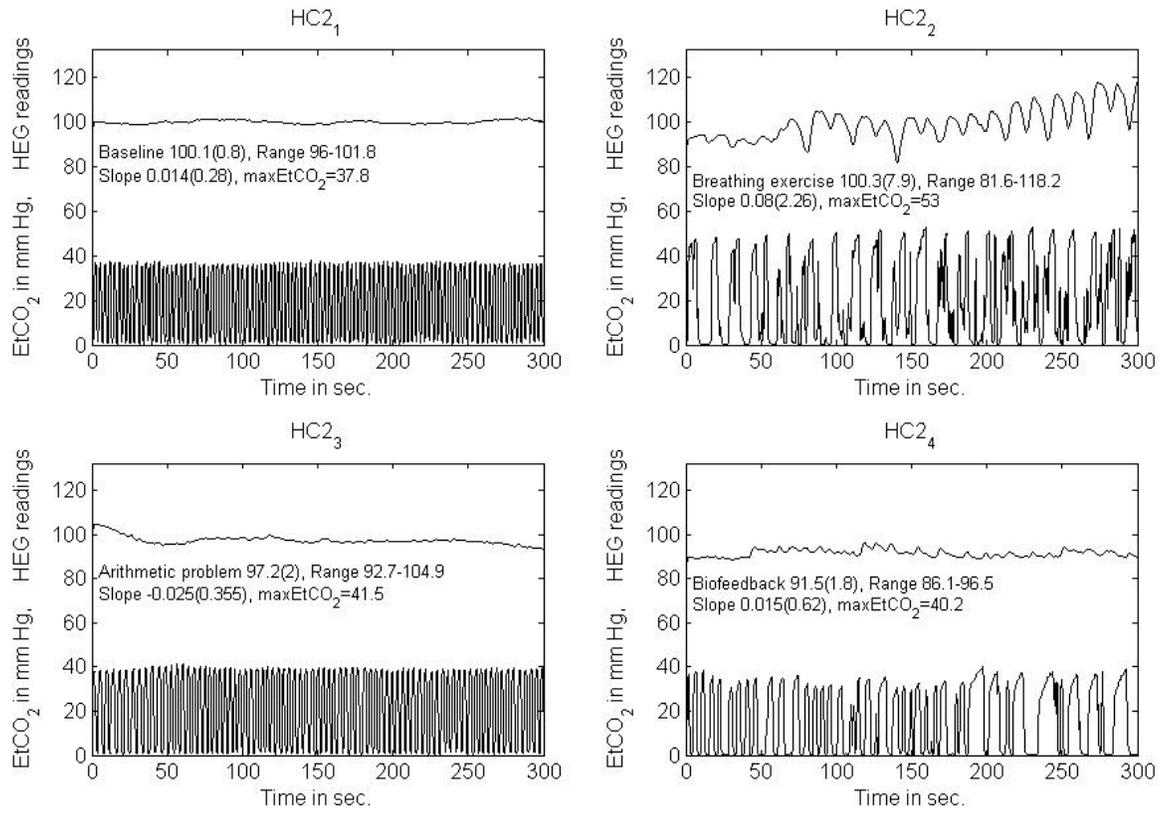

**Fig.4,** The HEG readings and the capnometer displays (EtCO$_2$) are shown for the 4 cases: baseline (subplot HC2$_1$), breathing exercise (subplot HC2$_2$), arithmetic problem (subplot HC2$_3$) and biofeedback (subplot HC2$_4$). The upper curves are the HEG readings, the lower curves the CO$_2$ values (at exhale).

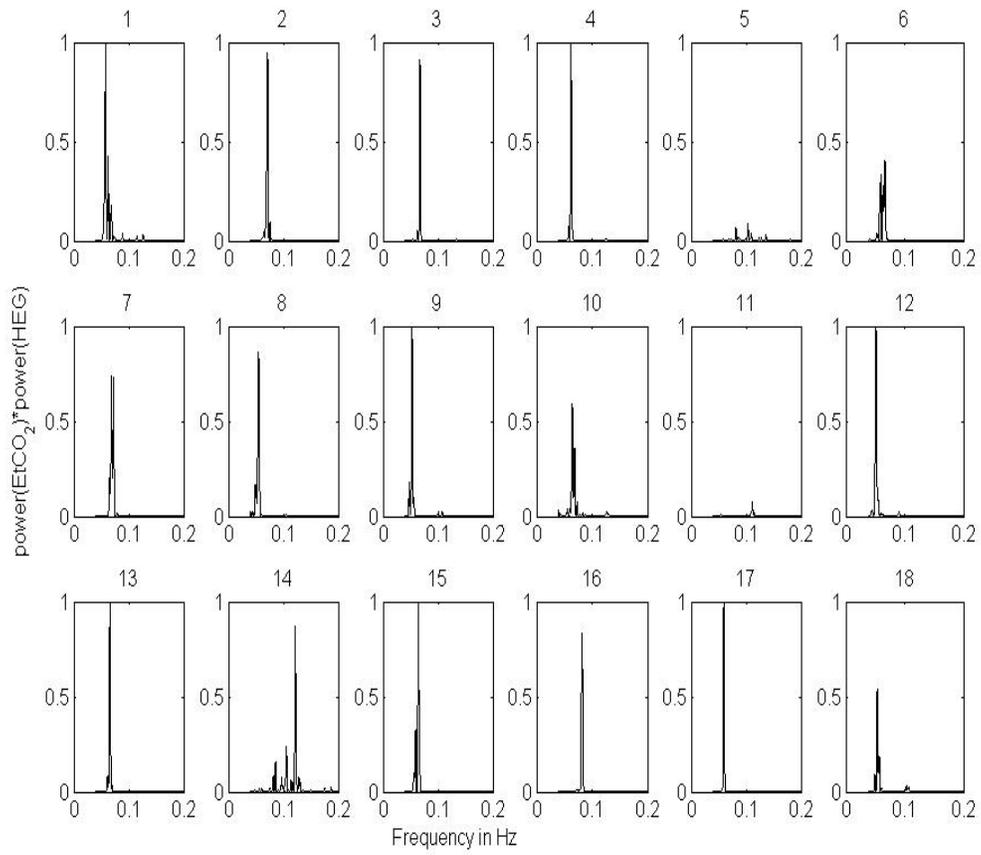

Figure 5. The correlation between the power spectra of the EtCO$_2$ periodic pattern and the corresponding HEG periodic pattern is depicted by their multiplication. The power spectra are normalized to unity. Maximal correlation is achieved when the multiplication is equal to 1.

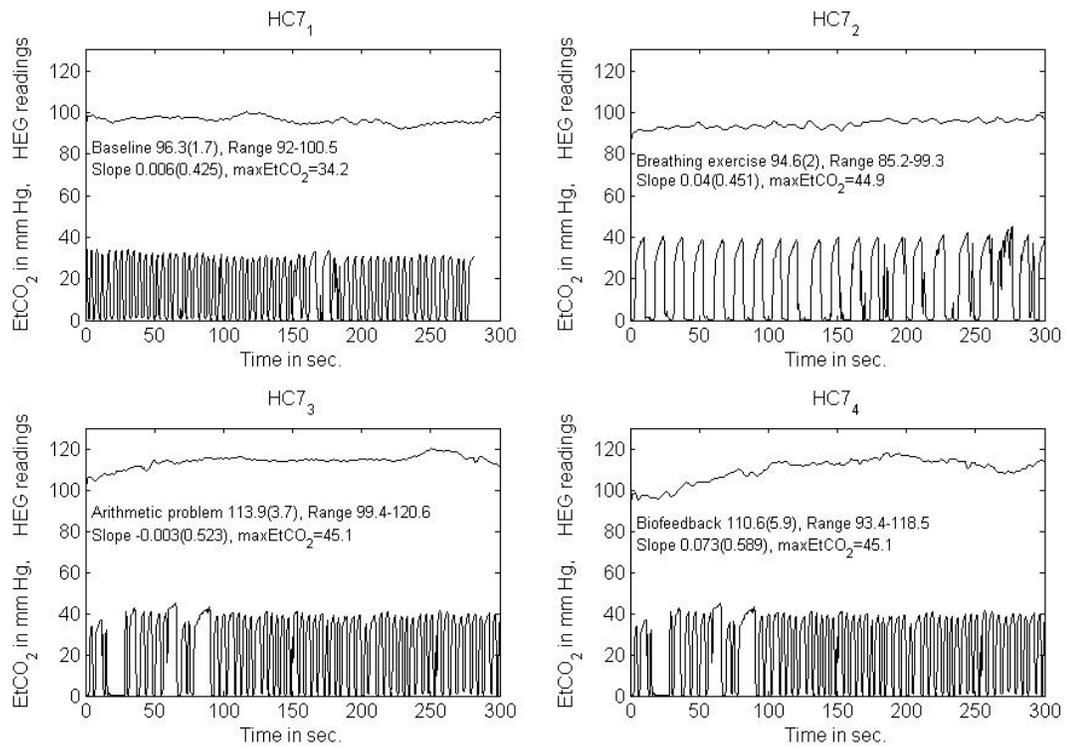

Fig. 6. Same as in Fig. 4 for participant No. 7.

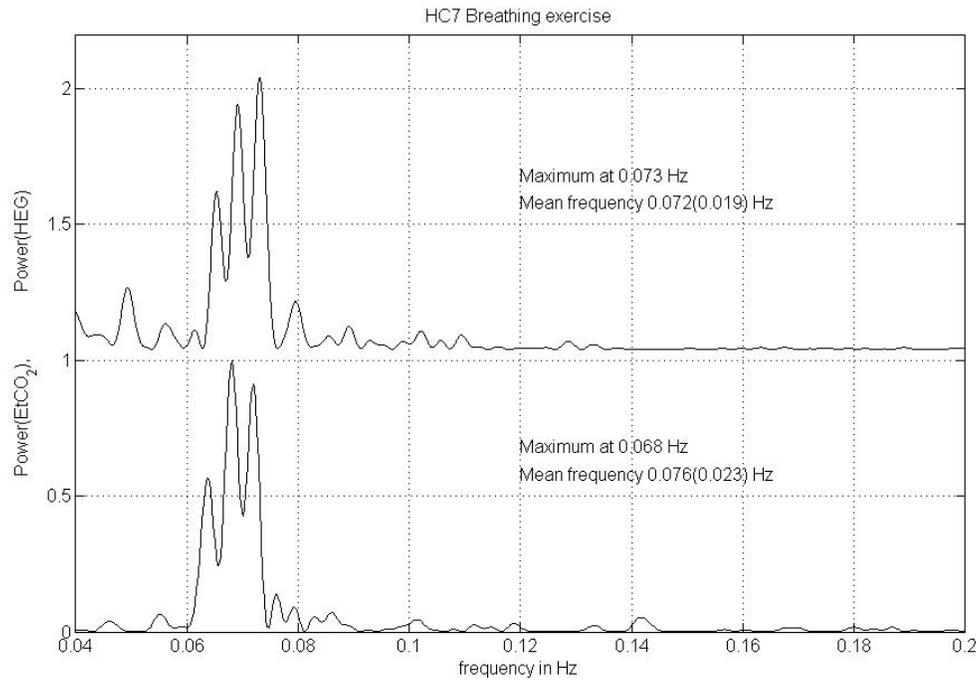

Fig. 7. Shown are the breathing exercise power spectra of the EtCO$_2$ and HEG patterns of participant No. 7. The similarity is very impressive, indicating that the periodicity is the same.

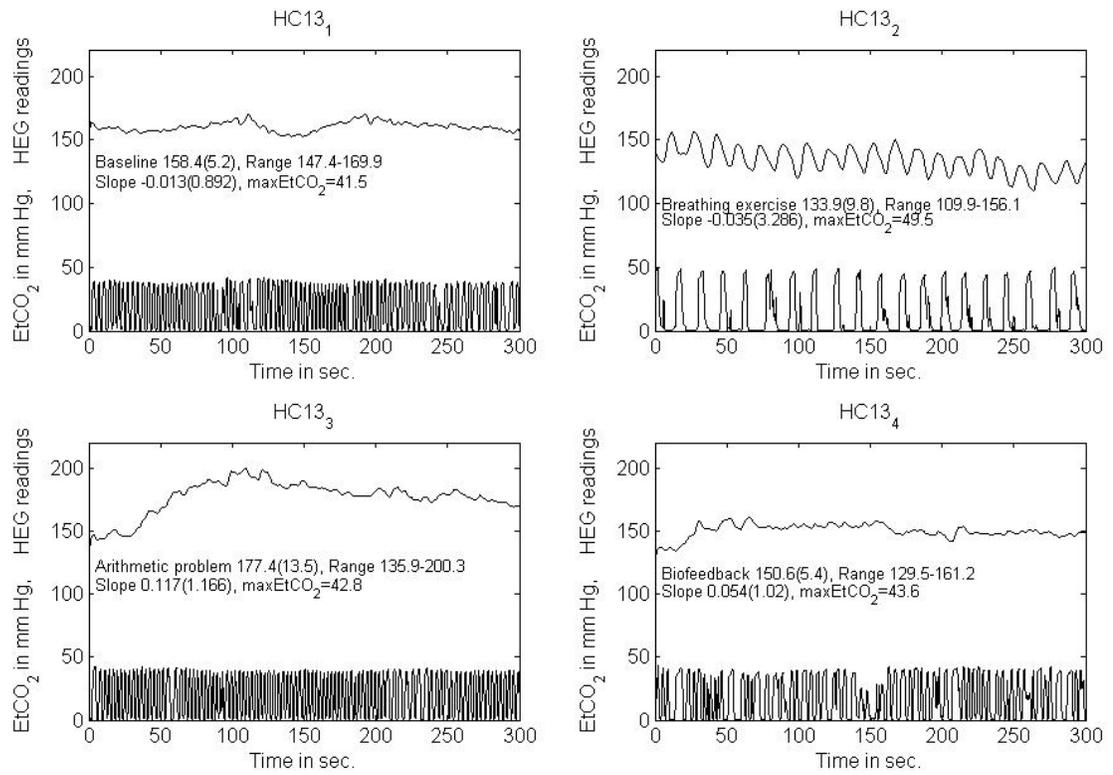

Fig. 8. The same as Fig. 4 for participant No. 13.

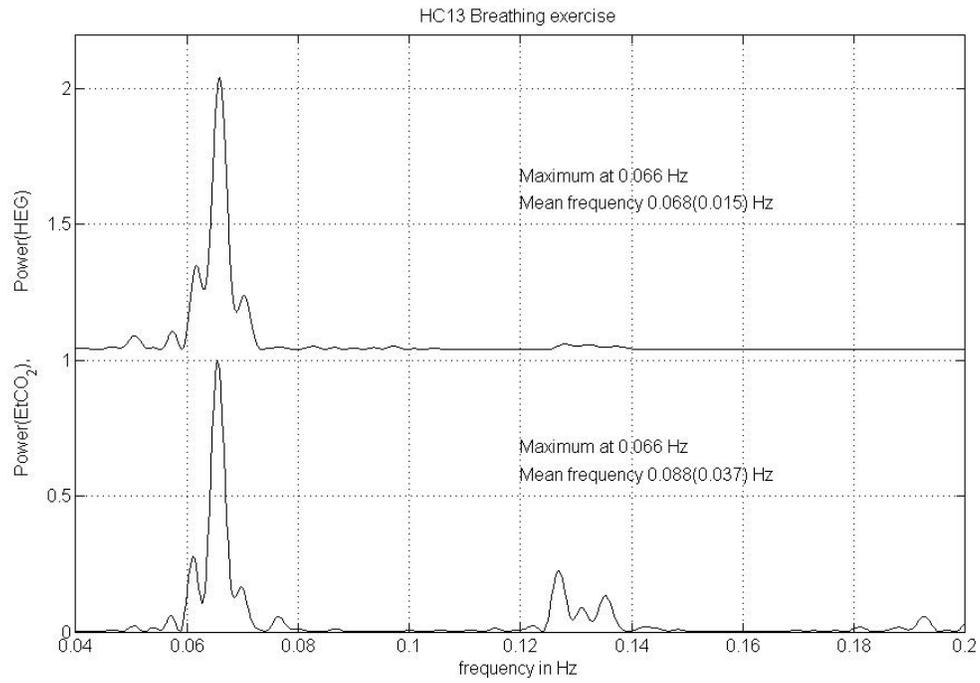

Fig. 9. The same as Fig. 7 for participant No. 13

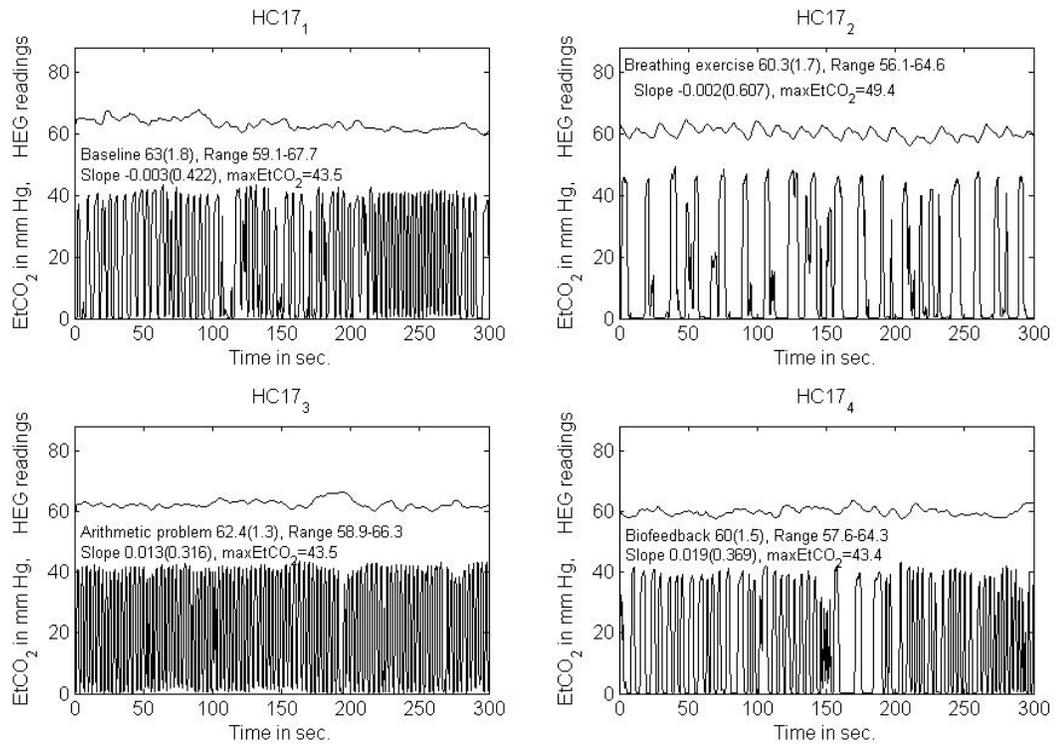

Fig. 10. The same as Fig. 4 for participant No.17.

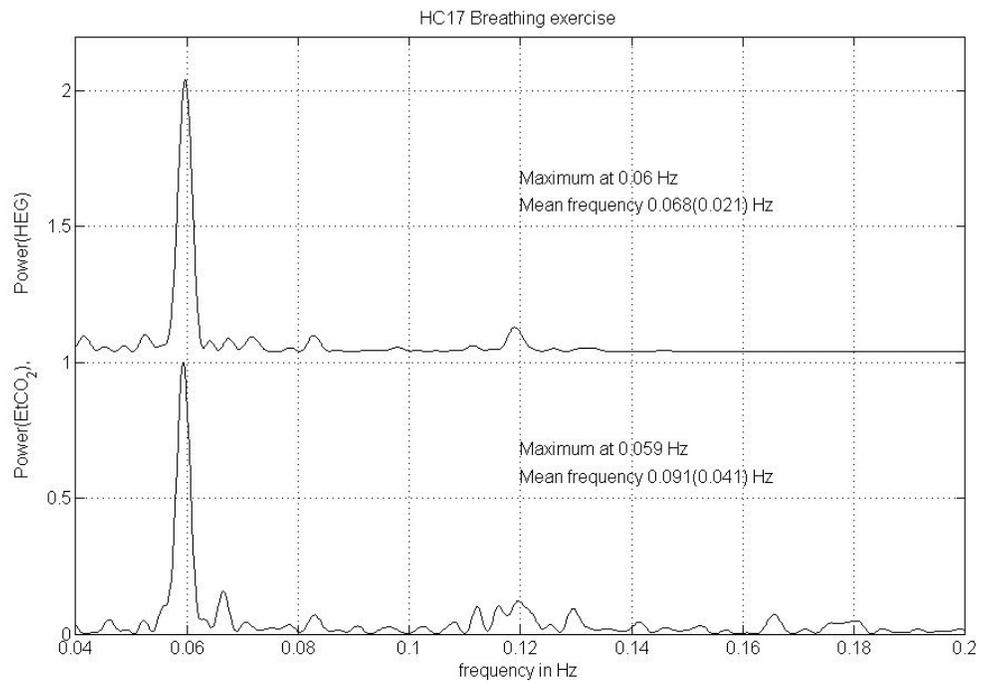

Fig. 11. The same as Fig.7 for participant No. 17.